\documentclass[12pt]{iopart}

\usepackage{citesort}
\usepackage{pdfsync}

\usepackage{graphicx}
\usepackage[caption=false]{subfig}
\usepackage{caption}
\usepackage{epstopdf}
\usepackage{dcolumn}
\usepackage{bm}
\usepackage{color}
\usepackage{float}
\usepackage[title, titletoc]{appendix}
\usepackage{setspace}
\usepackage{hyperref}
\usepackage{cite}
\usepackage{xr}
\usepackage{url}
\usepackage{booktabs}

\expandafter\let\csname equation*\endcsname\relax
\expandafter\let\csname endequation*\endcsname\relax
\usepackage{amsmath}
\usepackage{amssymb}

\begin{document}

\title{The Waiting-Time Distribution for Network Partitions in Cascading Failures in Power Networks}

\author{Long Huo$^1$, Xin Chen$^1$$^\dagger$}

\address{$^1$Center of Nanomaterials for Renewable Energy, State Key Laboratory of Electrical Insulation and Power Equipment, School of Electrical Engineering, Xi'an Jiaotong University, Xi'an 710054, Shaanxi, China}
\ead{xin.chen.nj@xjtu.edu.cn}
\vspace{10pt}
\begin{indented}
	\item[]Spet. 2021
\end{indented}

\begin{abstract}
Network redundancy is one of the spatial network structural properties critical to robustness against cascading failures in power networks. The waiting-time distributions for network partitions in cascading failures explain how the spatial network structures affect the cascading behaviors.  Two waiting time events associated with the first and largest network partitions are studied for cascading failures under different network redundancies. With the synthetic power networks, the waiting-time distributions of network partitions can be systematically analyzed for various network redundancies. Waiting-time distributions shift to the right accordingly when network redundancies increase. Meanwhile, the sizes of the largest partitions decrease while the numbers of them increase statistically. The realistic power networks of France, Texas, and Poland also show the same trend for waiting-time distributions. 
\end{abstract}

%
\vspace{2pc}
\noindent{\it Keywords}: cascading failure, waiting-time distribution, network redundancy, power flow dynamics
\\
%
\maketitle
%
%
\section{Introduction}\label{sec1}

The spatio-temporal dynamics of the complex power network attract a lot of interest due to the modernization of power systems and the increasing penetration of renewable energy sources\cite{rydin_gorjao_open_2020, 8481364, PhysRevLett.109.064101, PhysRevE.81.056106, nesti2018emergent, 8630092, zhao_spatio-temporal_2016,menck2014dead,lozano_role_2012}. Many empirical and theoretical studies have been conducted, such as cascading failures\cite{zhao_spatio-temporal_2016, nesti2018emergent, PhysRevLett.100.218701}, the synchronization in power networks\cite{PhysRevLett.109.064101, menck2014dead, lozano_role_2012}, the intermittent renewable energy fluctuations\cite{PhysRevE.81.056106}, and spatio-temporal distributions of power networks frequencies\cite{rydin_gorjao_open_2020, 8630092}. 
The relationship between the spatial network structure and the complex temporal dynamics is an important topic in different disciplines\cite{menck2014dead, martinez2018functional, barzel2013universality}. The robustness of power systems is still a long-term problem\cite{7435343, 4282033, 5275700}. Understanding the spatio-temporal properties of power networks is important to mitigate cascading failure and improve the system's robustness.\cite{tu2018optimal, wang2011robustness}

The origin of large but rare cascading failures triggered by small initial shocks is a phenomenon existing diversely\cite{valdez2020cascading}. Cascading failures in networked systems\cite{delvenne2015diffusion, zhao_spatio-temporal_2016} are universal for a wide range of real-world systems, from the blackout in power networks \cite{nesti2018emergent, rohden2016cascading, haes2019survey, nesti2020emergence}, informative block in communication networks \cite{wu2019dynamic, zhu2018modeling, ren2018stochastic} to the economic crisis in financial system \cite{lee2015forest}. In the anomalous diffusion, the waiting-time distribution often takes the form of the exponential power law\cite{rocha2013bursts, iribarren2009impact}.
Empirical evidence indicates that intricate temporal patterns of activity often characterize real-world networks, including a fat-tailed power-law waiting-time distribution, non-Markovianity, and non-stationarity\cite{min2009waiting, scholtes2014causality, horvath2014spreading}. The waiting-time distributions in the cascading failures of power networks are still unclear given the power flow physics\cite{yang2017small, yang2017vulnerability}. The structural properties of the complex networks significantly influenced the cascading dynamics\cite{po2017evolving} and therefore the waiting-time distribution\cite{tadic2005search, min2009waiting}. Existed work has shown that network redundancy is an essential structural property determining the robustness of power networks against cascading failures. \cite{plietzsch2016local}

In cascading failures, the small-scale initial failures possibly lead to a partition of the original integrated power network into disjoint components. The structures of power networks determine the waiting times for the partition events. Understanding network structures' impact on the waiting-time distribution can help reveal the spatial-temporal patterns of power networks. The partition events are very informational and critical for mitigating system failures, {\it{i.e.}}, the isolation control to prevent the failures from spreading over the entire networks\cite{young2019consequences, esmaeilian2016prevention}.

The rest of this paper are organized as follows. In Sec.~\ref{sec2}, we recall some necessary notations in graph theory. In Sec.~\ref{sec4}, we define the waiting events of network partition in cascading failures and discuss the corresponding waiting-time distributions for synthetic power networks with various network redundancies. We discuss the waiting-time distributions for the realistic power networks in Sec.~\ref{sec7}. The conclusions were drawn in Sec.~\ref{sec8}.

\section{Preliminaries and Notations}\label{sec2}
First, we recall some notations in graph theory, which will be used in the reminder of this paper. From the viewpoint of graph theory, the power network can be modeled as an unweighted and undirected graph $\mathcal{G}=\left(\mathcal{V},\mathcal{E},\mathbf{A} \right)$, where $\mathcal{V}=\left\{ 1,2,\cdots ,\mathrm{N} \right\}$ is the set of nodes including $N=\left| \mathcal{V} \right|$ different nodes, $\mathcal{E}\subset \mathcal{V}\times \mathcal{V}$ is the set of undirected edges including $M=\left| \mathcal{E} \right|$ different edges, and $\boldsymbol{A}\in \mathbb{R}^{N\times N}$ is the adjacency matrix. Each node in the graph $\mathcal{G}$ represents a generator bus or a load bus and each edge in the graph $\mathcal{G}$ represents a power transmission line in the power network. We use the terms of bus/node and line/edge interchangeably. An edge in $\mathcal{E}$ between node $i$ and $j$ is denoted as $l = (i,j)$.  The entries of $\boldsymbol{A}$ satisfy $a_{ij} = a_{ji} = 1$ for each edge $\left( i,j \right) \in \mathcal{E}$ and are zero otherwise. The incidence matrix $\boldsymbol{C}\in \mathbb{R}^{M\times N}$ of graph $\mathcal{G}$ is defined as,
\begin{equation}
	C_{il}=\begin{cases}
		1     ,node\,\,i\,\,is\,\,the\,\,source\,\,node\,\,of\,\,edge\,\,l\\
		-1   ,node\,\,i\,\,is\,\,the\,\,\sin k\,\,node\,\,of\,\,edge\,\,l\\
		0     ,otherwise\\
	\end{cases}\label{eq1}
\end{equation}
Since the connectivity of a power network is time-varying in cascading failures, the graph is time-dependent defined as $\mathcal{G}(t)=\left(\mathcal{V},\mathcal{E}(t),\mathbf{A}(t) \right)$. The edge set at time t is denoted as $\mathcal{E}(t)$. The number of edges that are functional at time t is $\overline{M}\left( t \right) $ where $0\leqslant \overline{M}\left( t \right) \leqslant M$ and the number of failed edges before time t is $M - \overline{M}\left( t \right)$.The adjacency matrix $\mathbf{A}(t)$ is time-dependent, in which $a_{ij}(t) = a_{ji}(t) = 1$ for each edge $\left( i,j \right) \in \mathcal{E}(t)$. The incidence matrix $\boldsymbol{C}(t)\in \mathbb{R}^{\overline{M}(t)\times N}$ is defined by Eq. \ref{eq1} for each edge $\left( i,j \right) \in \mathcal{E}(t)$.

\section{Waiting-times Distributions under Various Network Redundancies}\label{sec4}
Network redundancy is one of the essential spatial structural properties to network cascading behaviors in networks. 
To understand how network redundancies influence the waiting-time distributions in cascading failures systematically, the synthetic power networks with desired network redundancy are generated with the Synthetic Power Network Model (SPNM) that is discussed in detail in~\ref{sec4-1}.  The cascading failure dynamics are modeled with the DC power flow. The propagation dynamics of cascading failures on power networks are discussed in~\ref{sec3}. 

In a cascading failure, a network partition can be treated as an event that occurs at the waiting time relative to the initial cascading failure. There are two events of interest to describe the spatio-temporal behavior of network partitions. One is the event of the first network partition. The other is the event of the first largest partition. With the graph theory notations, the initial power network at t=0 is fully-connected and defined as $\mathcal{G}\left( 0 \right) =\left\{ \mathcal{V},\mathcal{E}\left( 0 \right). \boldsymbol{A}\left( 0 \right) \right\}$. $\mathcal{S}\left( \cdot \right)$ denotes the size of the power network.
$\mathcal{S}\left( \mathcal{G}\left( 0 \right) \right) = N$ is the size of the original network at t=0. At the time t of the cascading failure, the power network consists of $d$ different isolated subnetworks, $i.e.$, $\mathcal{G}^i\left( t \right) =\left\{ \mathcal{V}^i,\mathcal{E}^i\left( t \right) ,\boldsymbol{A}^i\left( t \right) \right\}$, $i=1,2,\cdots ,d$, $d\geqslant 2$, where $\mathcal{V} =\mathcal{V}^1 \cup \mathcal{V}^2 \cup \cdots \cup \mathcal{V}^d$ and $\mathcal{E}^i\left( t \right) \cap \mathcal{E}^j\left( t \right) =\oslash$ for $\lor i\ne j=1,2,\cdots ,d$. The size of a network partition is measured with the size change in the largest connected component (LCC) \cite{huang2008understanding,buldyrev2010catastrophic} before and after the network partition at time t. LCC is the largest isolated subnetwork, $\mathcal{G}^{\max}\left( t \right) :=\max _{\mathcal{G}^i\left( t \right) ,i=1,2,\cdots ,d}\,\left[ \mathcal{S}\left( \mathcal{G}^1\left( t \right) \right) ,\mathcal{S}\left( \mathcal{G}^2\left( t \right) \right) ,\cdots ,\mathcal{S}\left( \mathcal{G}^d\left( t \right) \right) \right]$. The size change of LCC can be characterized with the network disruption $\mathcal{S}\left( \bar{\mathcal{G}}^{\max}\left( t \right) \right)$. The normalized size of LCC in terms of the initial power network, $\hat{\mathcal{S}}\left( \mathcal{G}^{\max}\left( t \right) \right) =\mathcal{S}\left( \mathcal{G}^{\max}\left( t \right) \right) /\mathcal{S}\left( \mathcal{G}\left( 0 \right) \right)$, is used to characterize the network partitions on an equal footing. As a result, the network disruption at time t is normalized as $\hat{\mathcal{S}}\left( \bar{\mathcal{G}}^{\max}\left( t \right) \right) =\mathcal{S}\left( \bar{\mathcal{G}}^{\max}\left( t \right) \right) /\mathcal{S}\left( \mathcal{G}\left( 0 \right) \right)$.

\begin{figure}[htb]
	\centering
	\subfloat {\includegraphics[width=1\textwidth]{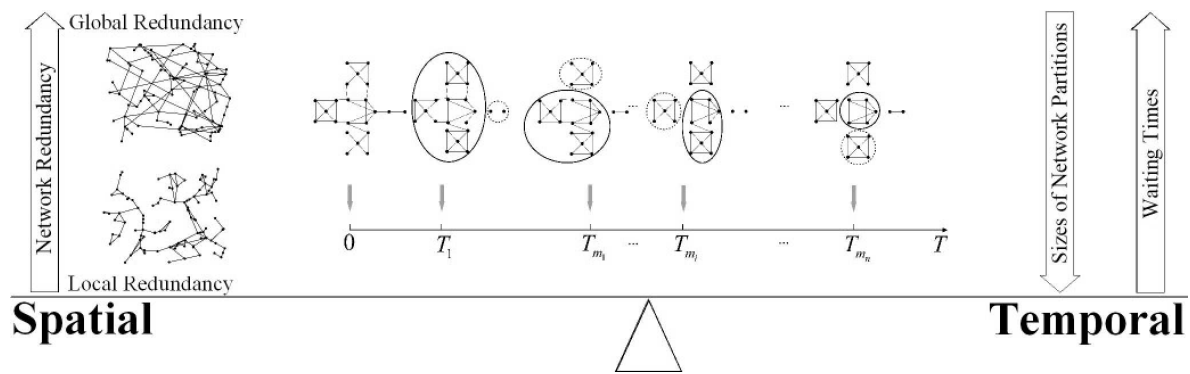}}\\ 
	\caption{The sequence of network partitions in the cascading failure is shown. The clusters in the solid-line circle are LCCs, and the clusters in the dashed-line circle are network disruptions at the waiting times of $T_1$ and $(T_{m_1}, \cdots, T_{m_n})$.}\label{fig1}
\end{figure}

The waiting time of the first partition is defined as $T_1:=\min _t\,\mathcal{S}\left( \mathcal{G}^{\max}\left( t \right) \right) <\mathcal{S}\left( \mathcal{G}^{\max}\left( 0 \right) \right)$. On the other hand, the waiting time of the first largest partition is defined as {\it{i.e.}} $T_{m_1}:= \min _t\max \mathcal{S}\left( \bar{\mathcal{G}}^{\max}\left( t \right) \right)$ with the event of the first largest disruption in the cascading failure. The largest network partitions may not be unique. Therefore, the waiting time $T_{m_1}$ can be extended to be a sequence of waiting times, $T_{m_i},\; i=\{1, \cdots, n\}$ and $T_{m_1} > \cdots > T_{m_i} > \cdots, >T_{m_n}$.
The corresponding largest network disruptions can be extended to be $\mathcal{S}\left( \bar{\mathcal{G}}^{\max}\left( T_{m_i} \right) \right), i=\{1,\cdots,n\}$. The waiting-time distributions of $T_1$, $T_{m_1}$ and $T_{m_i}$ provide temporal statistics for the cascading failure dynamics. Fig.~\ref{fig1} shows an illustration of a cascading failure with first network partition $T_1$ and the sequence of the largest network partitions $T_{m_i}$. The clusters in solid-line circles and dashed-line circles are LCCs and the clusters disconnected from LCCs in the cascading failure, respectively.

Network redundancy is a crucial network structural property for network robustness against cascading failures. In a cascading failure, redundancy allows the power network to remain connected by providing alternative paths. The edges in the network with the local form of network redundancy are more locally connected, increasing the possibility of the existence of triad-loop structures. The network redundancy can be measured with the transitivity and algebraic connectivity. The Laplacian matrix $\boldsymbol{L}\in \mathbb{R}^{N\times N}$ of graph $\mathcal{G}$ is $\boldsymbol{L}:=\mathrm{diag}\left( \sum\nolimits_{j=1}^N{a_{ij}} \right) -\boldsymbol{A}$. If $\mathcal{G}$
is connected, then $ker\left( \boldsymbol{C} \right) =ker\left( \boldsymbol{L} \right) =span\left( \boldsymbol{1}_N \right)$, where $\boldsymbol{1}_N$ is the vector with all one entries. The smallest eigenvalue of $\boldsymbol{L}$ is zero and all N - 1 remaining non-zero eigenvalues are strictly positive. The second-smallest eigenvalue is called the algebraic connectivity of graph $\mathcal{G}$, denoted as $\lambda _2\left( \boldsymbol{L} \right)$. The transitivity of $\mathcal{G}$ is defined as the fraction of closed-loop triads in terms of triads as: 
\begin{equation}
	\varGamma (\boldsymbol{A})=\frac{\sum_{i,j,k}{a_{ij}a_{jk}a_{ki}}}{\sum_{i,j,k,i\ne j}{a_{jk}a_{ki}}} \label{eq2}
\end{equation}
where $a_{ij}$, $a_{jk}$ and $a_{ki}$ are the entries in the adjacency matrix $\boldsymbol{A}$. The transitivity $\varGamma (\boldsymbol{A})$ in Eq.\ref{eq2} is used to measure local redundancy while the algebraic connectivity $\lambda _2\left( \boldsymbol{L} \right)$ is used to measure global redundancy. The more local network redundancy, ther larger transitivity $\varGamma (\boldsymbol{A})$.  However, the more global global network redundancy, the larger algebraic connectivity $\lambda _2\left( \boldsymbol{L} \right)$. Power networks with global redundancies tends to have a higher probability of finding long-range connections in networks. The competition between local and global redundancies is the structural trade-off between resilience against cascading failures and synchronization stability\cite{plietzsch2016local}. The network redundancies of synthetic power networks generated with the SPNM model are quantified with the redundancy parameters $r$. The synthetic power network with the smaller redundancy parameter r has more local redundancy. To study the influence of network redundancy on the waiting-time distributions, we consider three cases of synthetic power networks whose $r$ parameters are $10^{-2}$, $10^{0}$, and $10^{2}$ corresponding to the local, intermediate, and global redundancies. For each redundancy parameter $r$, 500 random synthetic power networks with 500 nodes are generated . The cascading failure is simulated according to the cascading failure dynamics. There are 25,0000 sample configurations where 500 initial failures sampled for each network.  

\begin{figure}[H]
	\centering
	\subfloat[$T_1$ Waiting-time Distribution]{\includegraphics[width=0.38\textwidth]{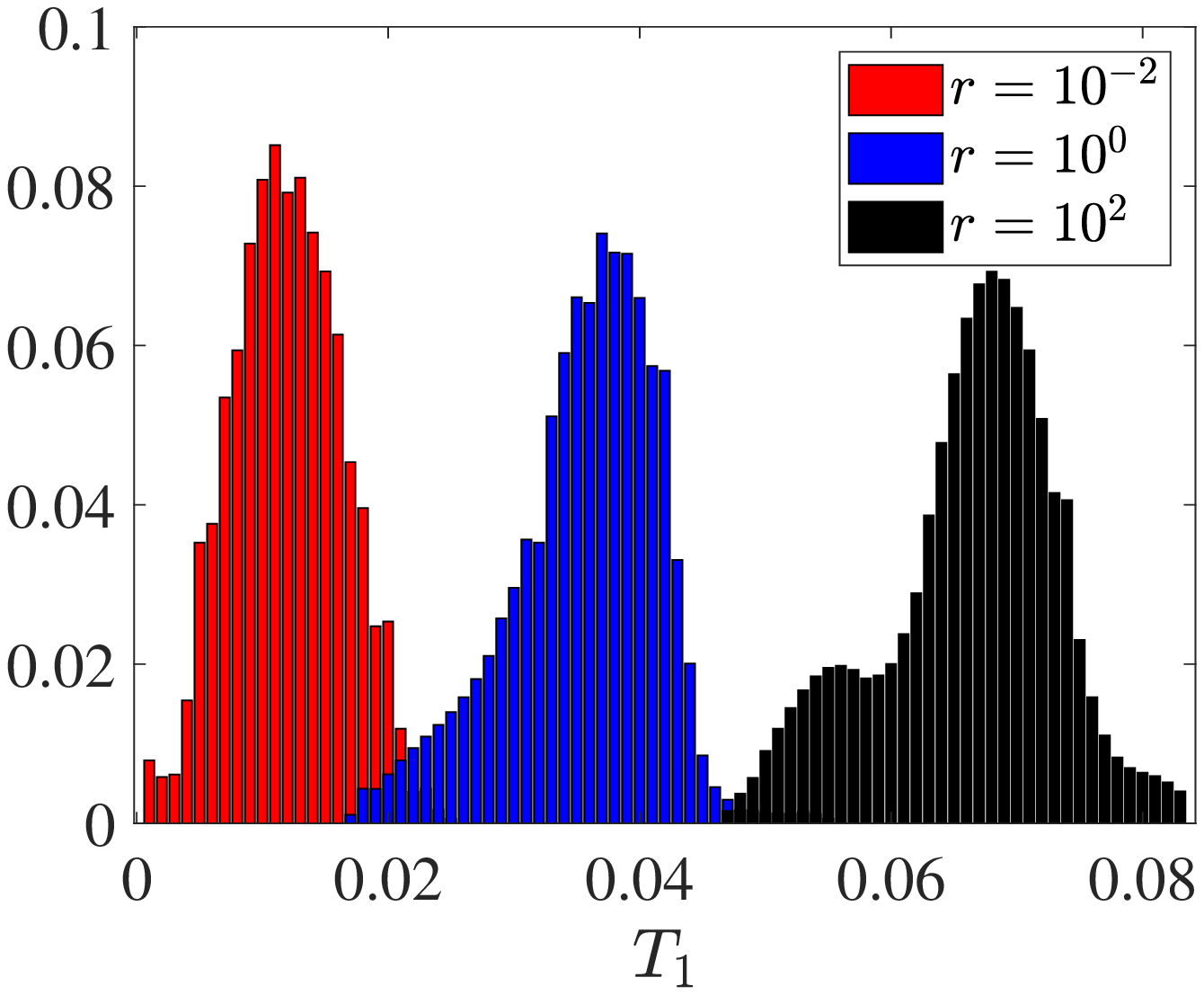}\label{fig3a}}
	\hspace{.3in}
	\subfloat[Cut-edge Ratio]{\includegraphics[width=0.38\textwidth]{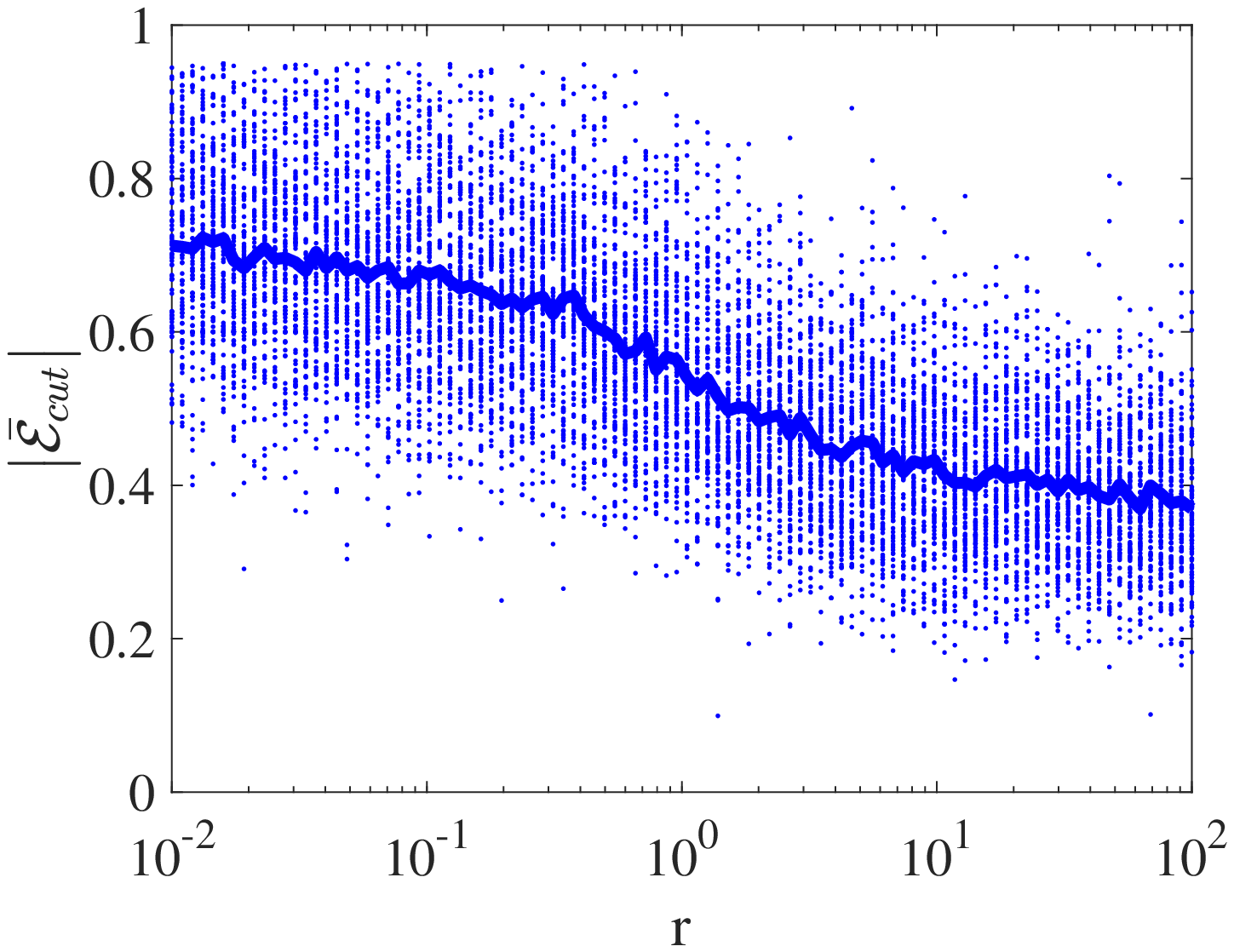}\label{fig3b}}
	\caption{(a) the $T_1$ waiting-time distribution of the first partition in the cascading failure and (b) the cut-edge ratio $\left| \bar{\mathcal{E}}_{cut} \right|$ $v.s.$ r. The parameter r is ranged from $10^{-2}$ to $10^{2}$. For each r, 500 networks with N=500 nodes are generated by SPNM. The blue dots are cut-edge ratio of each network and the solid line in blue is the average cut-edge ratio. }\label{fig3}
\end{figure}

Network redundancies strongly affect cascading failures. The $T_1$ waiting-time distribution has unimodal distribution or bimodal distribution as shown in Fig.~\ref{fig3a} for the local ($r=10^{-2}$), intermediate ($r=10^{0}$) and global ($r=10^{2}$) redundancies, respectively. When the global redundancy increases, the $T_1$ waiting-time distribution shifts to the right, {\it{i.e.}}, the waiting time $T_1$ are larger in the cascading failure. On average, the first partitions happen much later for the global redundancy. A cut edge is a bridge in graph theory, whose deletion must increase the number of connected components \cite{bollobas2013modern}. Cut edges don't exist in any closed-loop in connected graphs. Therefore, the cut-edge ratio plays a determining role in $T_1$ in power networks. The cut-edge ratio is defined as $\left| \bar{\mathcal{E}}_{cut} \right|=\left| \mathcal{E}_{cut}\left( 0 \right) \right|/\left| \mathcal{E}\left( 0 \right) \right|$, where $\left| \mathcal{E}_{cut}\left( 0 \right) \right|$ is the number of cut edges and $\left| \mathcal{E}\left( 0 \right) \right|$ is the number of all edges in the network. Fig.~\ref{fig3b} shows the cut-edge ratio $\left| \bar{\mathcal{E}}_{cut} \right|$ decreases with the increasing r. Networks are generated by SPNM with r ranged from $10^{-2}$ to $10^{2}$. The blue dots are cut-edge ratio of each networks and the solid line in blue is the average cut-edge ratio for each redundancy parameter r. Fig.~\ref{fig3b} indicates that the power networks with the global redundancy ($r=10^{2}$) have the smallest cut-edge ratio, and therefore, the network partition will happen least likely among the synthetic power networks with the local ($r=10^{-2}$), intermediate ($r=10^{0}$) and global ($r=10^{2}$) redundancies. As a result, the first network partition will happen much later.

\begin{figure}[H]
	\centering	
	\subfloat[Normalized Size of LCC]{\includegraphics[width=0.38\textwidth]{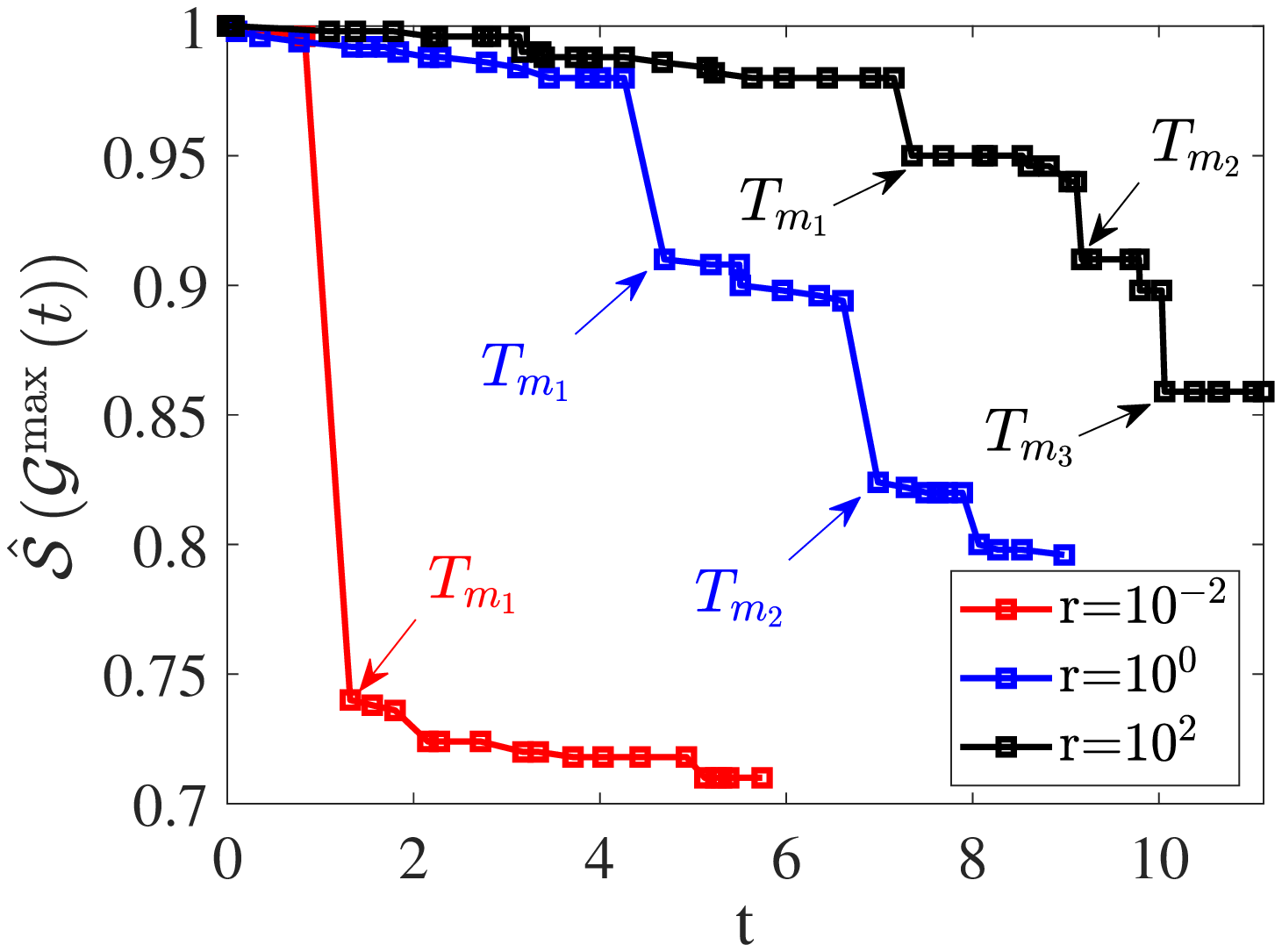}\label{fig4a}}
	\hspace{.3in}
	\subfloat[$T_{m_1}$ Waiting-time Distribution]{\includegraphics[width=0.38\textwidth]{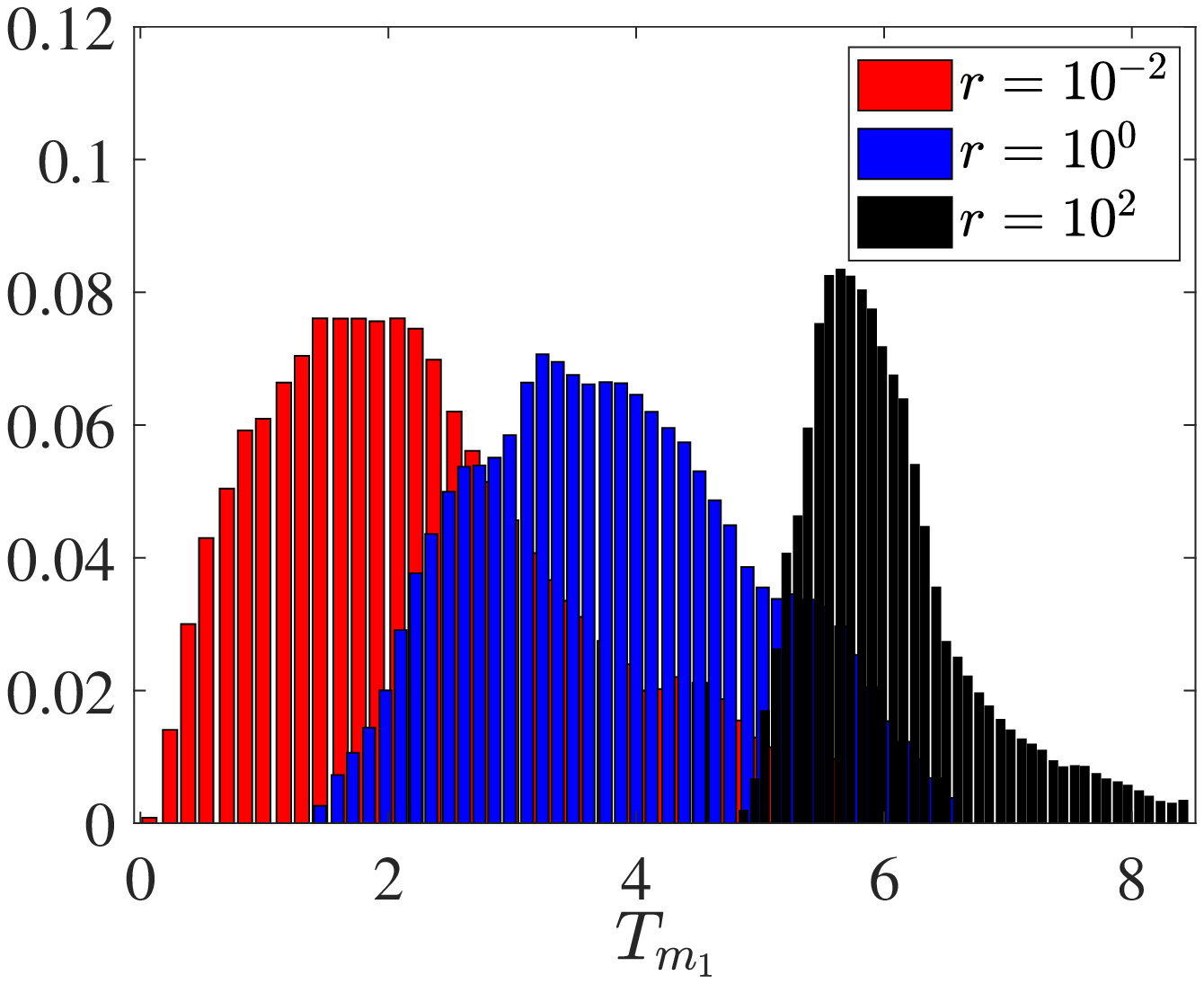}\label{fig4b}}\\
	\subfloat[$T_{m_i}$ Waiting-time Distribution]{\includegraphics[width=0.38\textwidth]{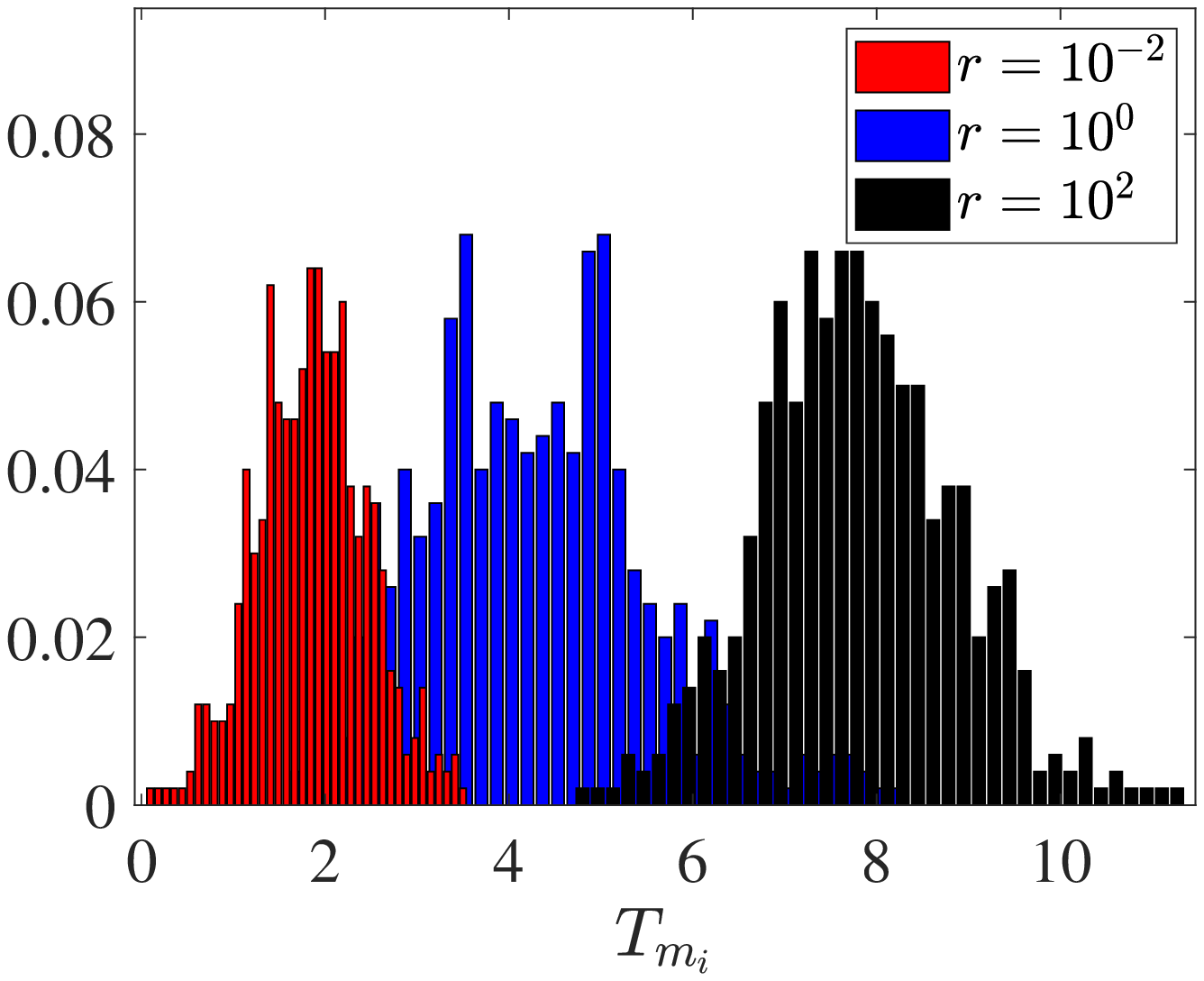}\label{fig4c}}
	\hspace{.3in}	
	\subfloat [Largest Disruption Distribution]{\includegraphics[width=0.38\textwidth]{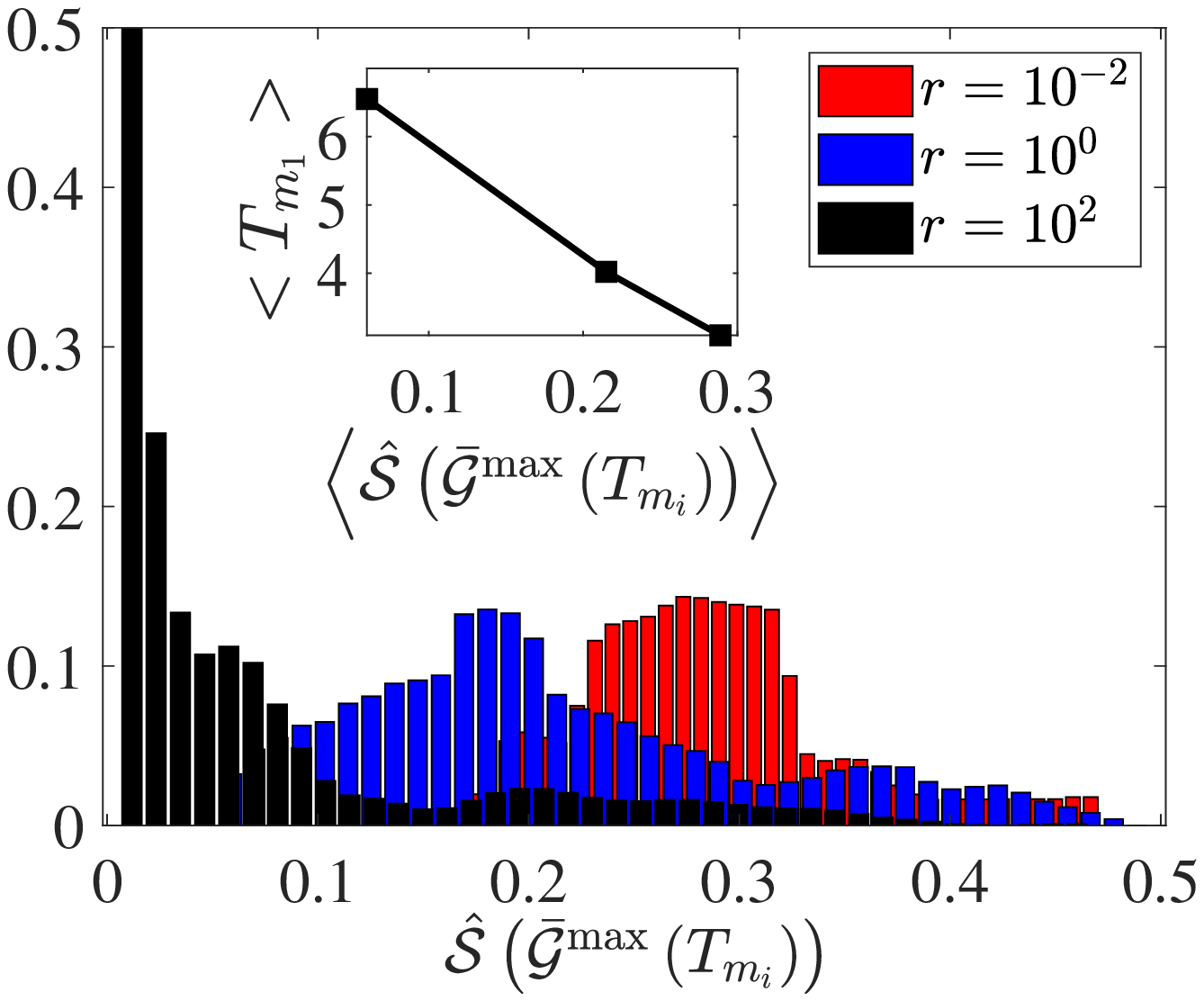}\label{fig4d}}	
	\caption{(a) shows an example of the normalized size of LCC $\hat{\mathcal{S}}\left( \mathcal{G}^{\max}\left( t \right) \right)$ for the networks with local ($r=10^{-2}$), intermediate ($r=10^{0}$) and global ($r=10^{2}$) redundancies during cascading failures. The red, blue and black dot lines correspond to the cascading failures happened in networks in Fig. \ref{fig2a} to Fig. \ref{fig2c}, respectively. (b) shows the $T_{m_1}$ waiting-time distribution of the first largest network partition. (c) shows the $T_{m_i}$ distribution of all the largest network partitions. (d) shows the distributions of the normalized largest network disruptions $\hat{\mathcal{S}}\left( \bar{\mathcal{G}}^{\max}\left( T_{m_i} \right) \right)$. The embedding plot in (d) shows the mean of largest disruptions $v.s.$ the mean of $T_{m_1}$. In (b) to (c), 500 synthetic power networks with N=500 nodes are generated by SPNM for local ($r=10^{-2}$), intermediate ($r=10^{0}$) and global ($r=10^{2}$) redundancies.}\label{fig4}
\end{figure}

Fig.~\ref{fig4a} shows the cascading failure in terms of the normalized size of LCC, $\hat{\mathcal{S}}\left( \mathcal{G}^{\max}\left( t \right) \right)$ for three typical synthetic power networks with local, intermediate and global redundancies shown in Fig. \ref{fig2a} to Fig. \ref{fig2c}. Among the three synthetic power networks, the one with local redundancy has the largest network disruption before and after the largest network partitions at $T_{m_i}$ and then the ones with intermediate and global redundancies. Since the synthetic power networks with global redundancy are more well connected, it is hard to break into different large clusters. 
On the other hand, the synthetic power networks with global redundancy have more network partitions than the other two. 
In summary, the size of the largest network disruptions decrease and the number of the largest partition events at $T_{m_i}$ increase for the synthetic power networks from the local redundancy to the global redundancy. 

The $T_{m_1}$ waiting-time distribution has unimodal distribution as shown in Fig.~\ref{fig4b} for the first largest partition events. For the events of the multiple largest network partitions, the $T_{m_i}$ waiting-time distribution has unimodal distribution for the local and global redundancies, as shown in Fig.~\ref{fig4c}. However, the $T_{m_i}$ waiting-time distribution has bimodal distribution for the intermediate redundancy since it has mixed local and global forms of network redundancy. Due to the structural difference of local and global redundancies, $T_{m_1}$ waiting-time distributions are different. For the local redundancy, the $T_{m_1}$ event happens earlier than the one for the global redundancy. 
Both $T_{m_1}$ and $T_{m_i}$ waiting-time distributions shift to the right as the redundancy parameter $r$ increases. 

The largest disruption distributions are shown in Fig.~\ref{fig4d}. The distributions are nearly unimodal distributions for the local and intermediate redundancies. For the global redundancy, the largest disruption distribution is more Pareto-like heavy-tailed. The average sizes of the largest disruptions decrease with network redundancies. Fig.~\ref{fig4d} shows the average waiting time $T_{m_1}$ , $<T_{m_1}>$ is negatively correlated with the average size of the largest disruption $<\hat{\mathcal{S}}\left( \bar{\mathcal{G}}^{\max}\left( t \right) \right)>$. 

The power networks with global redundancy tend to have the larger $T_{m_1}$ waiting times. The mean value of the $T_{m_1}$ waiting time $<T_{m_1}>$ increases gradually with the redundancy parameters. On average, the power networks with a global form of redundancy have a larger average waiting time $<T_{m_1}>$, and the corresponding largest network disruptions are smaller than those in the power networks with a local form of redundancy. In addition, the proportion of $T_{m_1}$ in $T_{m_i}$ also dramatically changes with network redundancy. For the local, intermediate and global redundancies, the proportion of $T_{m_1}$ in $T_{m_i}$ are $99.8\%$, $58.3\%$ and $43.4\%$ respectively. Clearly, for the local redundancy, almost all the largest partitions happened only once. For the global redundancy, the power networks tend to have the multiple largest partition more likely.

\section{Waiting-time Distribution in Realistic Power Networks}\label{sec7}
It is worthwhile examining the waiting-time distributions for realistic power networks. We choose the French, Texas, and Polish power networks, which have different network redundancies. The network topologies and electrical parameters of the three power networks can be found in MATPOWER toolbox\cite{5491276}. To compare the three realistic power networks on equal footing, synthetic power networks are used as a reference to qualify the network redundancy of the three realistic power networks. We define the relative transitivity and relative algebraic connectivity as $\varGamma (\boldsymbol{A})/\varGamma^* (\boldsymbol{A})$ and $\lambda _2\left(\boldsymbol{L} \right)/\lambda _2^*\left(\boldsymbol{L} \right)$, where $\varGamma (\boldsymbol{A})$ and $\lambda _2\left(\boldsymbol{L} \right)$ are the transitivities and algebraic connectivities of the three realistic power networks, $\varGamma^* (\boldsymbol{A})$ and $\lambda _2^*\left(\boldsymbol{L} \right)$ are transitivities and algebraic connectivities of the corresponding synthetic power networks generated by SPNM with the same sizes as the three realistic power networks. For each of the three realistic power networks, we generated 500 synthetic power networks with $r=10^{0}$ (intermediate redundancy). $\varGamma^* (\boldsymbol{A})$ and $\lambda _2^*\left(\boldsymbol{L} \right)$ are the averages of the transitivities and algebraic connectivities of the 500 synthetic power networks. If the ratio $\varGamma (\boldsymbol{A})/\varGamma^* (\boldsymbol{A}) > 1$ and $\lambda _2\left(\boldsymbol{L} \right)/\lambda _2^*\left(\boldsymbol{L} \right) < 1$, then the corresponding realistic power network is more local and vice versa. The network size $N$, average degree $\bar{k}$, relative transitivities, and algebraic connectivities of French, Texas, and Polish power networks are shown in Table~\ref{tab1}. It is found that the relative transitivities, $\varGamma (\boldsymbol{A})/\varGamma^* (\boldsymbol{A})$ are all larger than one, and the relative algebraic connectivities, $\lambda _2\left(\boldsymbol{L} \right)/\lambda _2^*\left(\boldsymbol{L} \right)$ all smaller than one for the three power networks. All the three realistic power networks have the local form of redundancy since the local redundancy means less building cost for the power network planning, and the redundancies decrease from the French to Texas to Polish power networks in terms of the relative transitivity and algebraic connectivity.

\begin{table}[H]
	\centering
	\caption{\label{tab1} Network size, average degree, transitivity and algebraic connectivity of the French, Texas and Polish power networks}
	\begin{tabular}{lclclcl}
\toprule & $N$ & $\bar{k}$ & $\varGamma (\boldsymbol{A})$ & $\lambda _2\left(\boldsymbol{L} \right)$ & $\varGamma (\boldsymbol{A})/\varGamma^* (\boldsymbol{A})$ & $\lambda _2\left(\boldsymbol{L} \right)/\lambda _2^*\left(\boldsymbol{L} \right)$\\ 
\midrule French Power Network & 2868 & 2.42 & 0.0370 & 0.0025 & 1.754 & 0.245\\   
Texas Power Network & 2000 & 2.67 & 0.0286 & 0.0028 & 1.356 & 0.275\\   
Polish Power Network & 2736 & 2.55 &0.0213 & 0.0035 & 1.010 & 0.343\\ 
\bottomrule   
	\end{tabular}  
\end{table}

\begin{figure}[H]
	\centering	
	\subfloat[$T_{1}$ Waiting-time Distributions]{\includegraphics[width=0.38\textwidth]{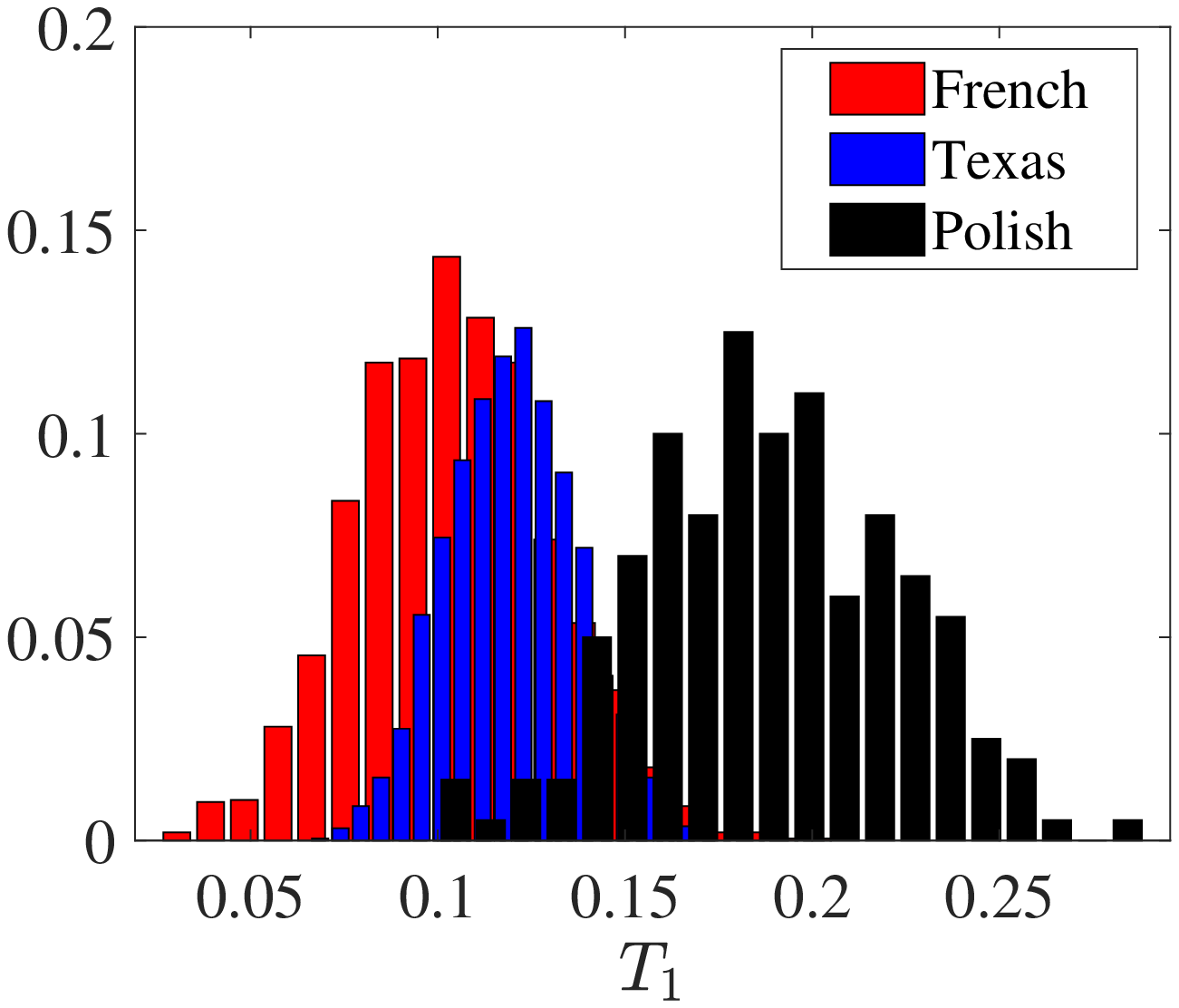}\label{fig6a}}
	\hspace{.3in}
	\subfloat[$T_{m_1}$ Waiting-time Distributions]{\includegraphics[width=0.38\textwidth]{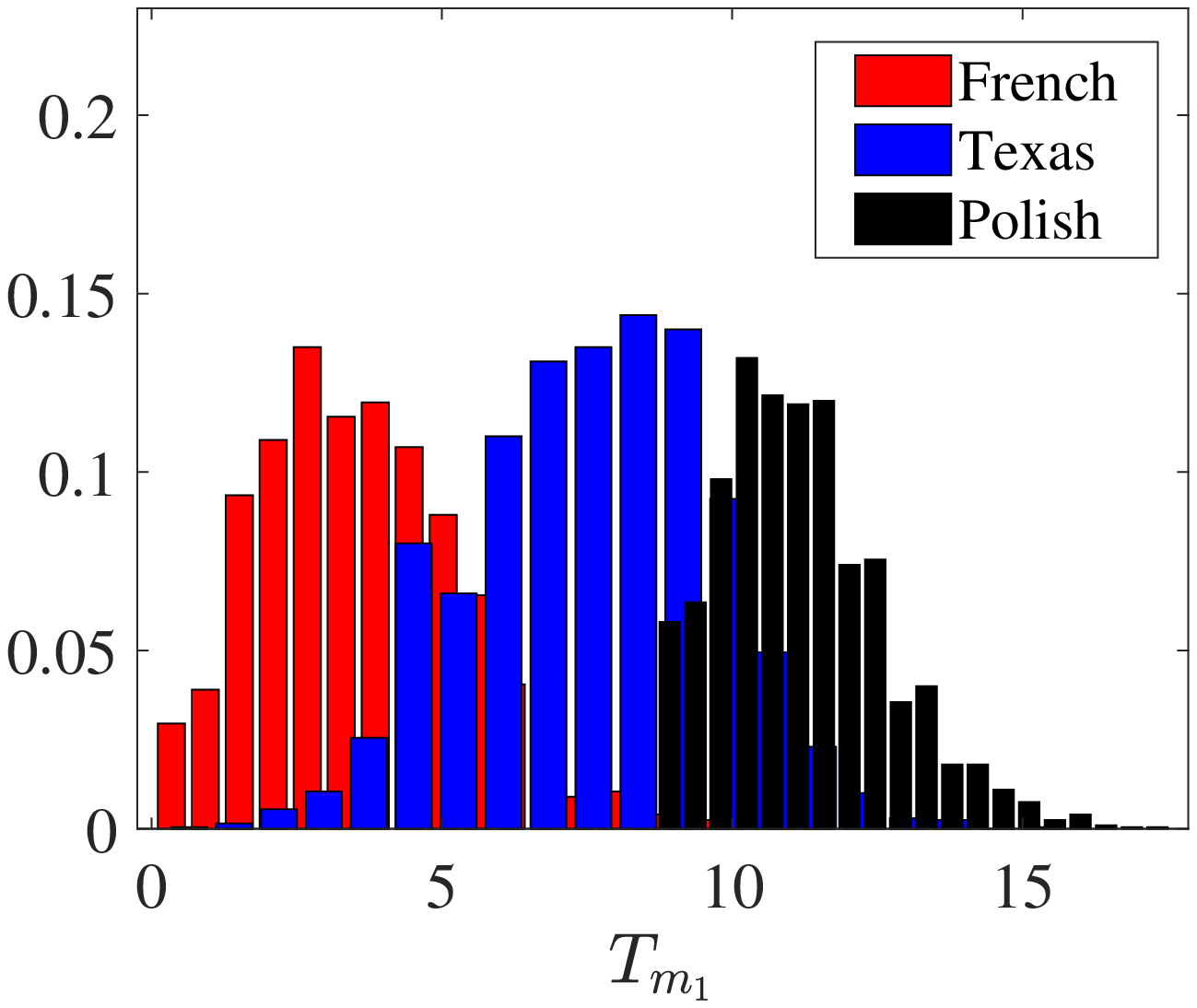}\label{fig6b}}
	\hspace{.3in}	
	\subfloat [Largest Disruption Distributions] {\includegraphics[width=0.38\textwidth]{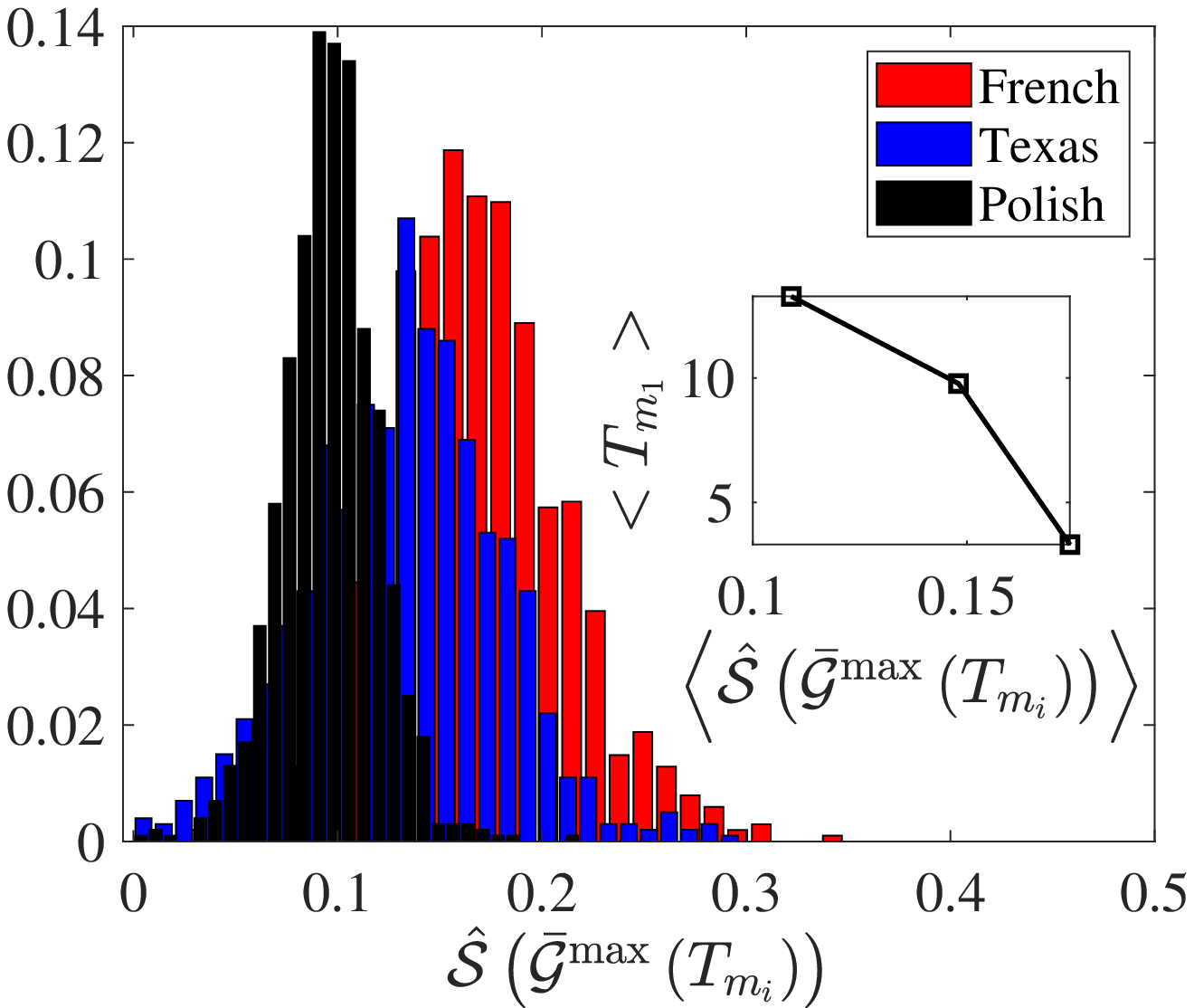}\label{fig6c}}	
	\caption{The $T_1$ and $T_{m_1}$ waiting-time distribution for the French, Texas  and Polish power networks.  (a) shows the $T_{1}$ waiting-time distribution. (b) shows the $T_{m_1}$ waiting-time distribution. (c) shows the largest disruption distribution, $\hat{\mathcal{S}}\left( \bar{\mathcal{G}}^{\max}\left( t \right) \right)$. The embedding plot in (d) shows the average sizes of the largest disruptions $v.s.$ the average $T_{m_1}$ waiting time.}\label{fig6}
\end{figure}

Fig. \ref{fig6} shows the $T_1$ and $T_{m_1}$ waiting-time distributions for the French, Texas and Polish power networks. In Fig. \ref{fig6a}, the $T_1$ waiting-time distributions shift to the right from the French to Texas to Polish power networks. 
The cut-edge ratio of French, Texas, and Polish power networks is 0.54, 0.14, and 0.13, respectively. 
For the French, Texas and Polish power networks, the largest disruption distributions of $\hat{\mathcal{S}}\left( \bar{\mathcal{G}}^{\max}\left( t \right) \right)$ are all nearly unimodal distributions and shift to the right from the Polish to Texas to French power networks as shown in Fig.\ref{fig6c}
The average sizes of the largest disruptions, $<\hat{\mathcal{S}}\left( \bar{\mathcal{G}}^{\max}\left( t \right) \right)>$ decreases with the average $T_{m_1}$ waiting time, $<T_{m_1}>$. In Summary, the three realistic power networks show the same trend for the waiting-time distributions with network redundancies as synthetic power networks in Section~\ref{sec4}. 

\section{Discussion and Concluding Remarks}\label{sec8}
The waiting-time distribution is strongly affected by the network structures of power networks. With the synthetic power networks tuned with the redundancy parameter r, the waiting-time distributions for the local, intermediate and global redundancies are studied systematically. The cascading failures are strongly affected by the network redundancy. For cascading failures in power networks, the $T_1$ waiting time of the first network partition and the $T_{m_1}$ waiting time of the first largest network partition will shift to the right from the local redundancy to the global redundancy. The largest network disruptions at $T_{m_i}$ will get smaller for larger global redundancies. Besides, the largest network partition in power networks may not be unique. The multiple largest network partitions happen more likely for global redundancy, and the sizes are much smaller. 

The waiting-time distributions can help understand how to tackle and mitigate the propagation of cascading failures in power networks. The network partitions in the cascading failures lead to the disconnection of different regions of the power networks and deteriorate the stability of frequency synchronization \cite{7435343}. Furthermore, building long-range transmission lines connecting different regions will increase the global network redundancy and delay large-scale power network partitions. On the other hand, a priori knowledge of network partitions and the temporal statistical properties will help the power network operators to prevent the propagation of failures efficiently and strategically.  It is of great interest to study how the higher-order structures influence the cascading failures in power networks in future work.

\section*{Acknowledgements}
This work was supported in part by the National Natural Science Foundation of China (Grant No.21773182 (B030103)) and the HPC Platform, Xi'an Jiaotong University. 

\begin{appendices} 
\renewcommand{\thesection}{Appendix A}
\section{Synthetic Power Networks}\label{sec4-1}
Synthetic power networks can help studying how network redundancy influences the waiting-time distributions in cascading failures systematically. SPNM is developed to generate synthetic power networks with desired network redundancy. SPNM includes two parts. One is the random growth model used to build the network topology with desired redundancy\cite{schultz2014random}. The other is the assignment of electrical parameters for the power network, including bus type, transmission line impedance, and the power of generation and load according to the empirical distributions of electrical parameters\cite{wang2016generating, wang2010generating, elyas2017statistical}. SPNM has the following three steps:
\begin{itemize}
	\item \textbf{Step 1:} The network topology of synthetic power network are generated by the random growth model\cite{schultz2014random}. In the random growth model, six parameters of $[N, N_0, p, q, s, r]$ are used to determine the topological properties, where $N$ is the total number of nodes in the power network, $N_0$  the initial number of nodes, $p$ the probability of generating an additional redundancy line attached at each new node added, $q$ the probability of generating further redundancy lines between existing nodes, $s$ the probability to split existing lines, and r is for local-vs-global redundancy trade-off. In the study, $r$ is the parameter to tune the network redundancy. The parameters $N=500$, $N_{0}=1$, $p=0.2$, $q=0.2$, $s=0$ are used in this paper for the random growth model to generate the topology of power network. The networks with more local form of redundancy have larger transitivity $\varGamma (\boldsymbol{A})$ while the networks with more global form of redundancy have larger algebraic connectivity $\lambda _2\left( \boldsymbol{L} \right)$. Fig. \ref{FIGlg} shows the transitivity and algebraic connectivity of the networks generated with 100 r uniformly selected in the range from $r=10^{-2}$ to $r=10^{2}$. With each r parameter, 500 networks are generated. The dots in blue and red are transitivity and algebraic connectivity of each network, respectively. The solid line in blue and red are the average value of transitivity and algebraic connectivity respectively. It can be found that the transitivity decreases with the increasing r and the algebraic connectivity increases with the increasing r, which indicates that the synthetic power networks with more local/global form of redundancy can be generated by the random growth model with smaller/larger r parameters.
	
	\begin{figure}[H]
		\centering
		\includegraphics[width=0.45\textwidth]{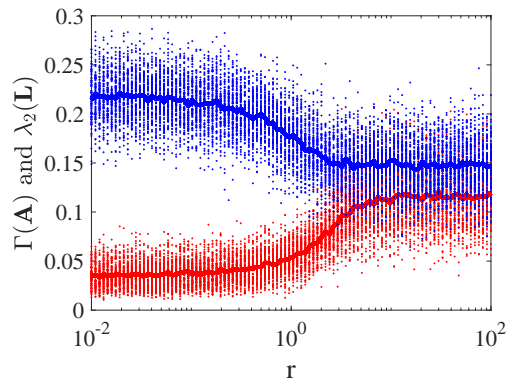}
		\caption{The transitivity $\varGamma (\boldsymbol{A})$ and the algebraic connectivity $\lambda _2\left( \boldsymbol{L} \right)$ of networks with r parameter ranged from $r=10^{-2}$ to $r=10^{2}$.  For each parameter $r$, 500 networks with 500 nodes are generated by the random growth model. The dots in blue and red are transitivity and algebraic connectivity of each network, respectively. The solid line in blue and red are the average value of transitivity and algebraic connectivity, respectively.}\label{FIGlg}
	\end{figure}
	
	\item \textbf{Step 2:} After the network topology is determined by step 1, all the nodes in the network need to be assigned with two node types: generation node, load node. The distribution of node types in real power system is not arbitrary, which can be measured by the bus type entropy\cite{wang2016generating}: 
	\begin{equation}
		E\left( \mathbb{T} \right) =-\sum_{k=1}^2{g_k\log g_k}-\sum_{k=1}^3{l_k\log l_k} \label{eq7}
	\end{equation}
	where $\mathbb{T}$ is a binary vector $\mathbb{T}=\left[ \mathbb{T}_i \right] _{N\times 1}$ of the bus types. $\mathbb{T}_i = 1$ if the $i$th node is the generation node and $\mathbb{T}_i = 0$ otherwise. The proportions of two types of nodes are $g_k$, {\it{i.e.}}, $g_1$ is the proportion of generation nodes and $g_2$ the proportion of load nodes. The proportions of three types of edges are $l_k$, {\it{i.e.}}, $l_1$ is the proportion of the generation-generation edges, $l_2$ the proportion of load-load edges, and $l_3$ the proportion of generation-load edges. The node type assignment could be transformed into a 0-1 programming problem:
	\begin{equation}
		\underset{\mathbb{T}}{arg}\,\,\min \left| E^*-E\left( \mathbb{T} \right) \right|\,\,    s.t.\lVert \mathbb{T} \rVert _1=\rho N \label{eq8}
	\end{equation}
	where the bus type entropy of a real power system, $E^*$ , is given as a reference. The objective is to obtain a binary vector $\mathbb{T}$ minimizing the difference between $E\left( \mathbb{T} \right)$ and $E^*$. $\lVert \cdot \rVert _1$ is the $L_1$ norm and $\lVert \mathbb{T} \rVert _1$ is the number of generation nodes. The parameter $\rho$ in the restraint is the expected ratio of generation nodes. A typical proportion of generation nodes\cite{wang2016generating} for real power system is 20\% - 40\%. The 0-1 programming in Eq.~\ref{eq8} can be solved by the genetic algorithm. As a classical swarm intelligence algorithm, the genetic algorithm\cite{whitley1994genetic} will initialize a population of individuals (which is also called "gene"). Each of the individuals is a vector representing the decision variables in the corresponding optimization problem, {\it{i.e.}}, each individual is a binary vector $\mathbb{T}$ in our case. Eq.~\ref{eq8} is the objective function and constraint. There will be a best individual whose objective is the minimum among the population at every iteration. The genetic algorithm finds an optimal solution successfully when the mean value of all individuals' objectives converges to the best individual's objective within the given maximum iterations. We have validated that to generate a synthetic power network with size N=500, the genetic algorithm will converge within 100 generations.
	
	\item \textbf{Step 3:} The electrical parameters of synthetic power network, {\it{i.e.}} the impedance of edges, are sampled from the probability density distribution (PDF) of the impedances in real power system. The PDF is the generalized Pareto distribution or double-Pareto log-normal distribution~\cite{wang2010generating}. According to the non-trivial correlation between the exponential generation and load distribution and the degrees of nodes, the algorithm~\cite{elyas2017statistical} is used to assign the generation and load to each node. 
\end{itemize}

\begin{figure}[H]
	\centering
	\subfloat[Local Redundancy($r=0.01$)]{\includegraphics[width=0.31\textwidth]{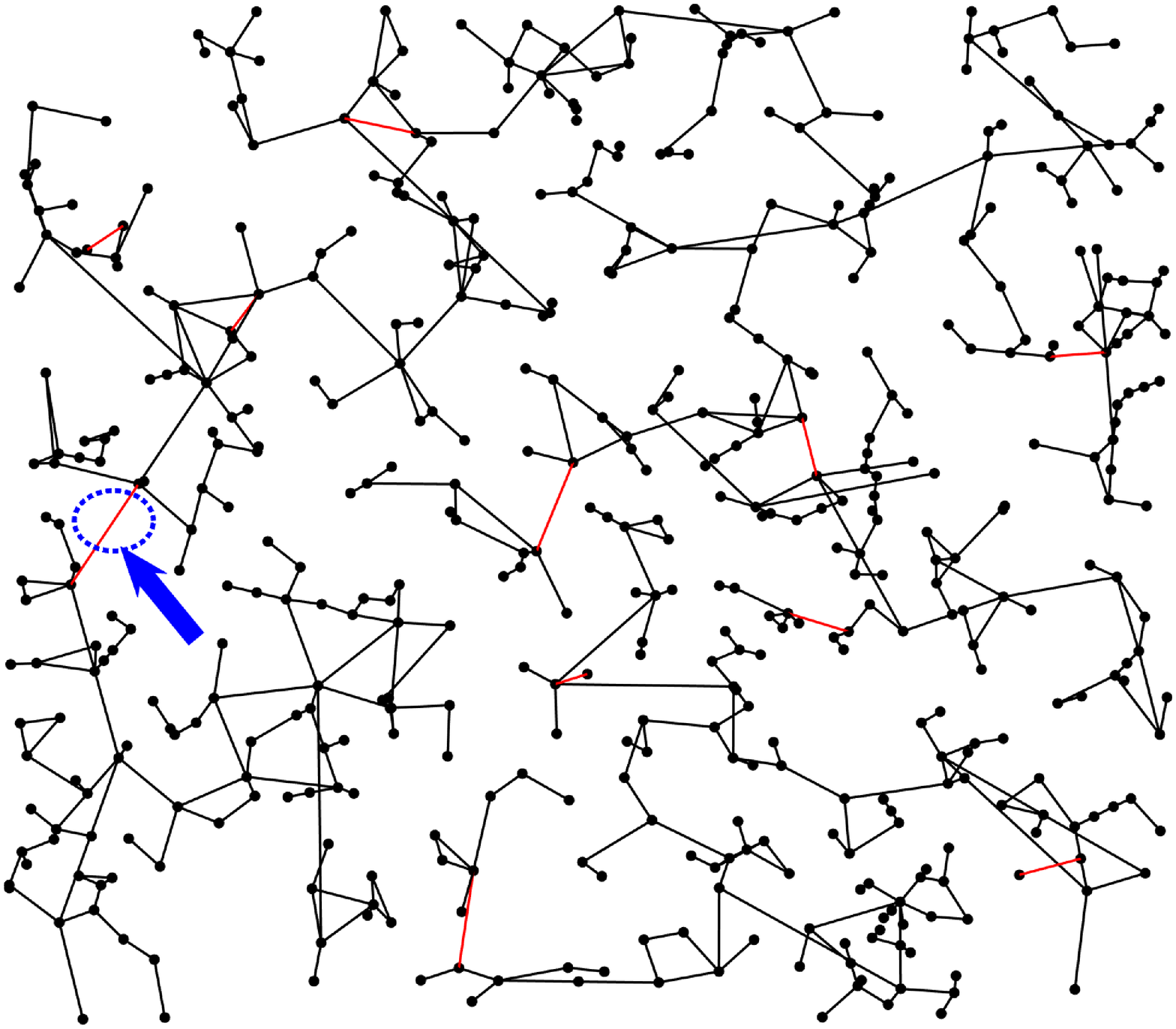}\label{fig2a}}
	\hspace{.1in}
	\subfloat[Intermediate Redundancy($r=1$)]{\includegraphics[width=0.31\textwidth]{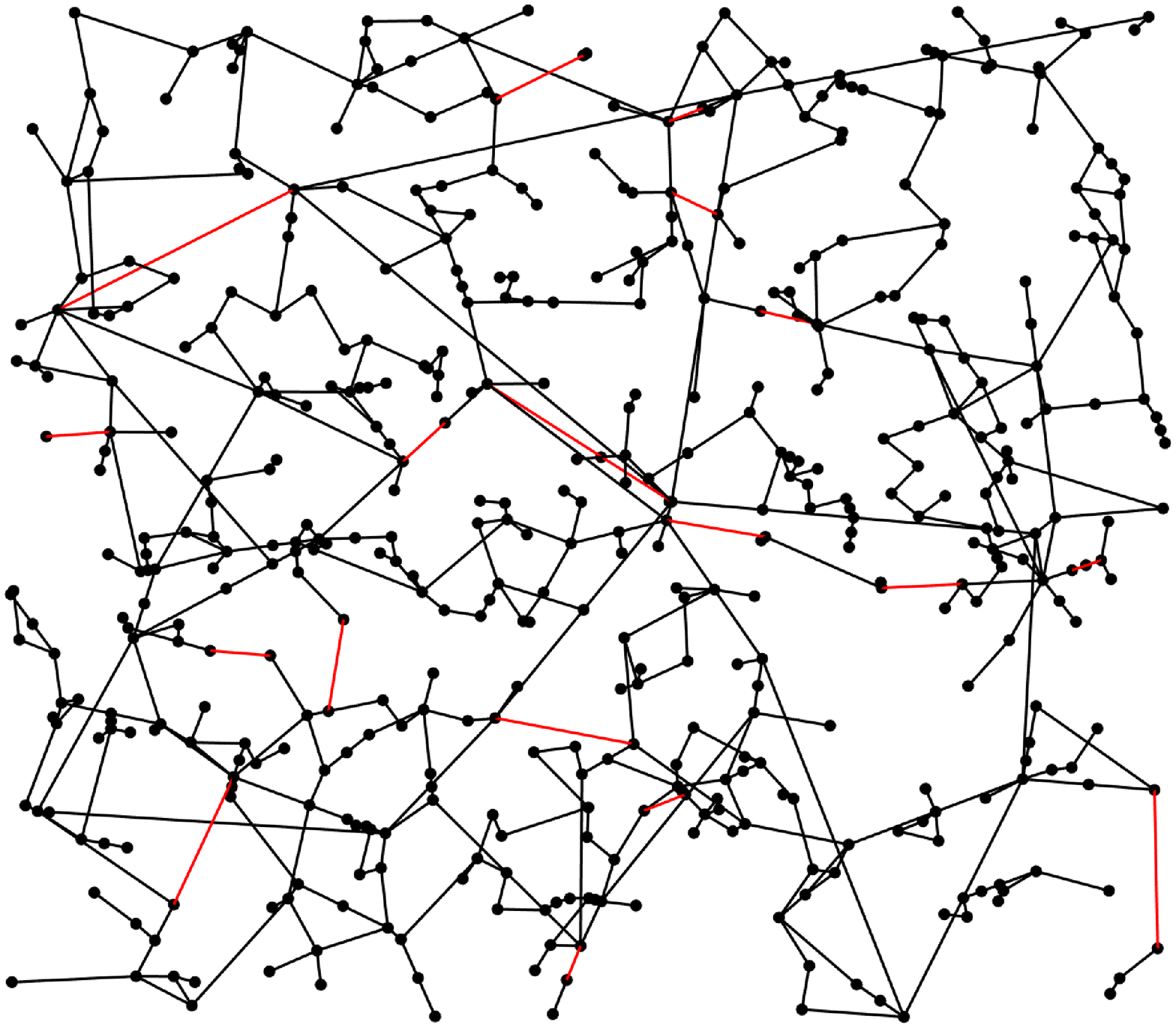}\label{fig2b}}
	\hspace{.1in}
	\subfloat[Global Redundancy ($r=100$)]{\includegraphics[width=0.31\textwidth]{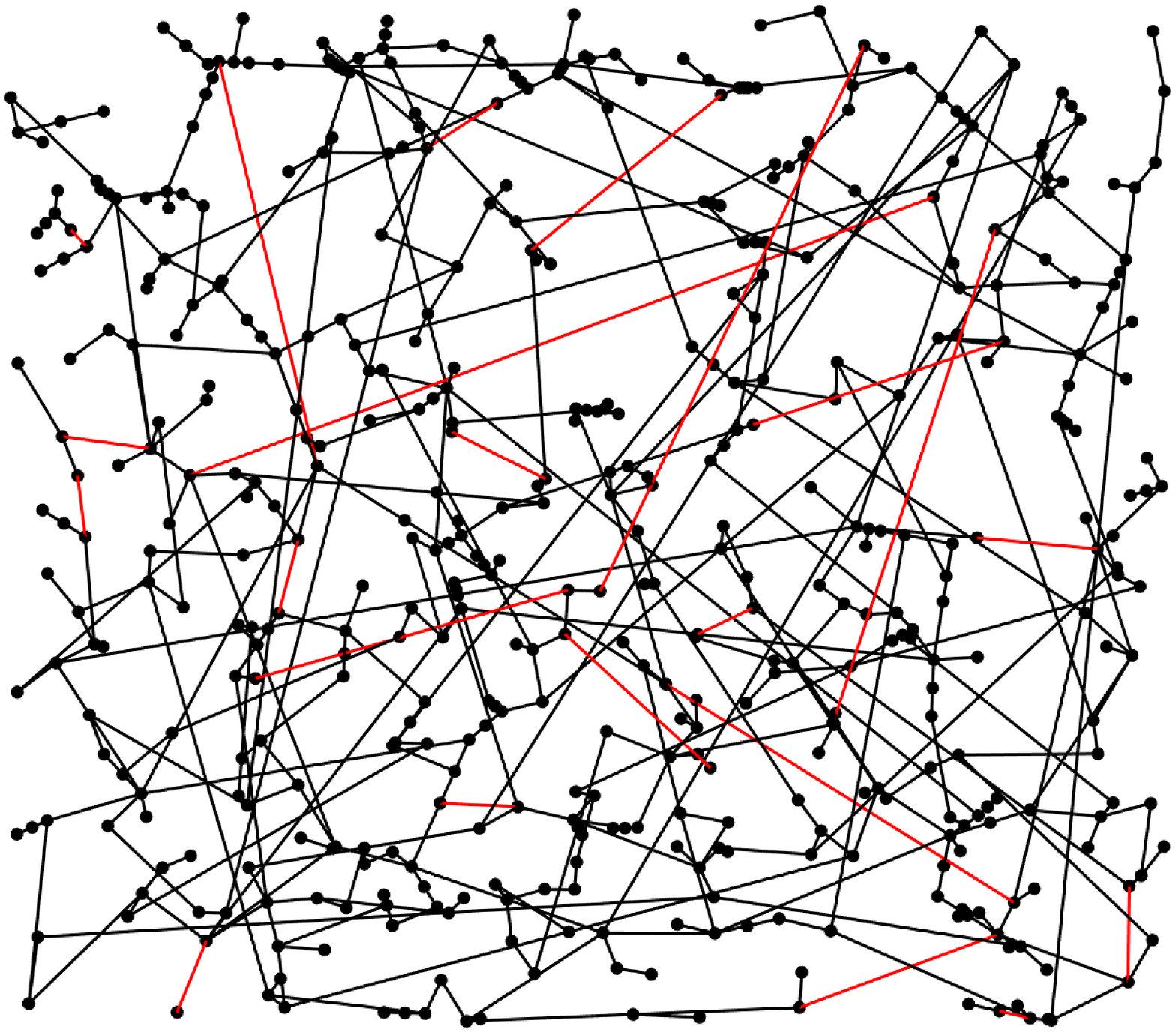}\label{fig2c}}	
	\caption{Three networks with local ($r=10^{-2}$), intermediate ($r=10^{0}$), and global ($r=10^{2}$) form of redundancies are presented in (a), (b) and (c) respectively. The dots in black are generator or load nodes. The edges in black are edges that are not failed and the edges in red are failed edges in the cascading failures. In (a), (b) and (c), the transitivity of network $\varGamma (\boldsymbol{A})$ is 0.234, 0.147 and 0.122  and the algebraic connectivity of network $\lambda _2\left( \boldsymbol{L} \right)$ is 0.041, 0.077 and 0.167.}\label{fig2}
\end{figure}

For typical power networks, the average degree is 2.8 roughly \cite{pagani2013power}. The synthetic power networks use the same average degree to enforce realistic power networks' structural constraints. Fig.~\ref{fig2} shows the three typical networks with the local redundancy ($r=10^{-2}(0.01)$), intermediate redundancy ($r=10^{0}(1)$), and global redundancy ($r=10^{2}(100)$) respectively. The three networks have the transitivity $\varGamma (\boldsymbol{A}) = 0.234, 0.147, 0.122$ and the algebraic connectivity $\lambda _2\left( \boldsymbol{L} \right) = 0.041, 0.077, 0.167$. The dots in black are generator or load nodes. The line in black are the edges that are not failed, and the ones in red are failed edges in the cascading failure. 

In Fig.~\ref{fig2a}, the network with the local redundancy clearly consists of the locally clustering structures and the connections between local clusters are sparse. For example, it is can be observed that only one inter-cluster edge \cite{brandes2005network} (indicated by the blue arrow and dashed circle) connects the cluster in the left corner and the remaining network. This edge usually has high betweenness centrality \cite{girvan2002community} and carries a large amount of power flow with a high overload risk. As a result, the largest partition happens when the cascading failure propagates to hit the edge. The network disruption is enormous. On the other hand, for the network with global redundancy, it has no clear locally clustering structures, and the network is densely connected with long-range edges as shown in Fig.~\ref{fig2c}. The largest partition results from a gradually accumulated failures and the network disruption is much smaller compared to the local redundancy. For the intermediate redundancy, Fig.~\ref{fig2b} shows there exists the mixing structure between the local and global redundancies. 

\end{appendices}

\begin{appendices} 
\renewcommand{\thesection}{Appendix B}
\section{Cascading Failure Dynamics}\label{sec3}
The quasi-steady state DC power flow is used to model the cascading failure dynamics\cite{9380543,yang2017small}. $\boldsymbol{F}(t)\in \mathbb{R}^{\overline{M}(t)\times 1}$ is the vector of the power flows on all edges $\overline{M}\left( t \right) $ in a power network $\mathcal{G}(t)$ at time t, where the element ${F}_l(t)$ denotes the power flow on edge $l\in \mathcal{E}(t)$. $\boldsymbol{P}(t)\in \mathbb{R}^{N\times 1}$ and $\boldsymbol{\theta}(t)\in \mathbb{R}^{N\times 1}$ are the vectors of the power injections and voltage angle phases on all nodes in the power network $\mathcal{G}(t)$ at time t, where the elements ${P}_i(t)$ and ${\theta}_i(t)$ denote the power injection and voltage angle phase at node $i\in \mathcal{V}$. The susceptance of edge $l$ in the power network is $B_l$ and we introduced the susceptance matrix $\boldsymbol{B}(t)\in \mathbb{R}^{\overline{M}(t)\times \overline{M}(t)}$ as $\boldsymbol{B}(t):=\mathrm{diag}\left( B_l \right) $, $l\in \mathcal{E}(t)$. The DC power flow can be described as:

\begin{equation}
	\boldsymbol{P}(t)=\boldsymbol{C}(t)\boldsymbol{F}(t) \label{eq3}
\end{equation}

\begin{equation}
	\boldsymbol{F}\left( t \right) =\boldsymbol{B}\left( t \right) \boldsymbol{C}^T\left( t \right) \boldsymbol{\theta }\left( t \right) \label{eq4} 	
\end{equation}
where $\left( \cdot \right) ^T$ is the matrix transpose operation. Eq. \ref{eq3} and \ref{eq4} basically are the Kirchhoff’s law and Ohm’s laws in power networks, respectively. For a balanced power injection vector $\boldsymbol{P}(t)$, $\sum\nolimits_{i\in \mathcal{V}}^{}{P_i\left( t \right)}=0$, the DC power flow in Eqs.~\ref{eq3} and ~\ref{eq4} has an unique solution $\left( \boldsymbol{F}\left( t \right) ,\boldsymbol{\theta }\left( t \right) \right)$ based on the reference voltage angle phase of the slack node in power network. The slack node is a generator bus whose power can be adjusted between zero and its maximum generation and it's voltage angle phase is fixed to a known value, $i.e.$, usually we set ${\theta}_i\left( t \right) =0$ if node $i$ is slack node. The DC power flow is widely used in electrical engineering and physics communities and is computationally efficient for simulation of cascading failures in large-scale power networks\cite{yang2017vulnerability, nesti2020emergence}. The DC power flow model is a good approximation to the AC high-voltage power transmission network. \cite{yang2017small}

Based on the DC power flow, the cascading failure dynamics can be modeled according to the following three steps:
\begin{itemize}
	\item \textbf{Step 1: initial failure}. The power network before cascading failure is fully connected without isolated sub-networks or node. It is really rare to have more than one initial failure simultaneously. Therefore, only one edge $l\in \mathcal{E}\left( 0 \right) $ will be randomly selected as the initial failure and removed from the original power network $\mathcal{G}(0)$. 
	
	\item \textbf{Step 2: power flow re-distribution}. After the initial failure, the power flow re-distribution needs to be calculated according to the DC power flow shown in Eq. \ref{eq3} and \ref{eq4} if the network remains fully-connected. The power network may break into several isolated sub-networks at time t, $i.e.$, the power network at time t $\mathcal{G}(t)$ consists of $d$ isolated sub-networks $\mathcal{G}^i\left( t \right) =\left\{ \mathcal{V}^i,\mathcal{E}^i\left( t \right) ,\boldsymbol{A}^i\left( t \right) \right\}$, $i=1,2,\cdots ,d$, $d \geqslant 2$, where  ${V}^i \cap {V}^j =\oslash$  and $\mathcal{E}^i\left( t \right) \cap \mathcal{E}^j\left( t \right) =\oslash$ for $i\ne j$. Although sub-networks change over the evolution of cascading failure, the total nodes $\mathcal{V}$ in the power network doesn't change as the sum of the nodes in each sub-network, $\mathcal{V} =\mathcal{V}^1 \cup \mathcal{V}^2 \cup \cdots \cup \mathcal{V}^d$. In each sub-network, the power consumption and generation need to be balanced first. In order to obtain the balanced power consumption and generation in the sub-network $\mathcal{G}^i(t)$, the generator node with largest generation capacity is selected as the slack node and the power of the slack node will be adjusted within the range between zero and its maximum generation to make sure $\sum\nolimits_{j\in \mathcal{V}^i}^{}{P_j\left( t \right) =0}$. If the power consumption and generation in the sub-network $\mathcal{G}^i(t)$ can't be balanced after the adjustment of the slack node, we uniformly scale down the power of all generator nodes (or load nodes) in the sub-network when $\sum\nolimits_{j\in \mathcal{V}^i}^{}{P_j\left( t \right) >0}$ (or $\sum\nolimits_{j\in \mathcal{V}^i}^{}{P_j\left( t \right) <0}$). Then the power flow re-distribution needs to be calculated independently for each sub-network. 
	
	\item \textbf{Step 3: edge removal}. Once the power flow re-distribution is done in each sub-network at time $t^{\prime}$, according to the temperature-evolution model\cite{yang2017small}, the heat starts to accumulate on each edge as described in Eq.~\ref{eq5}. The temperature of edge $l\in \mathcal{E}\left( t \right)$ at time t is given as:
\begin{equation}
	T_l\left( t \right) =e^{-ut}\left[ T_l\left( t^{\prime} \right) -T_e\left( \boldsymbol{F}_l\left( t^{\prime} \right) \right) \right] +T_e\left( \boldsymbol{F}_l\left( t^{\prime} \right) \right) \label{eq5}
\end{equation}
where $T_l\left( t^{\prime} \right)$ is the initial temperature of edge $l$ at time $t^{\prime}$, $T_e\left( \boldsymbol{F}_l\left( t^{\prime} \right) \right) =\frac{\rho}{\mu}\boldsymbol{F}_{l}^{2}\left( t^{\prime} \right) +T_a$ is the equilibrium temperature as $T_l\left( t \right)$ evolves to $t\rightarrow \infty$.  The constants $\rho$ and $\mu$ are determined by the electrical characteristics of the power transmission line, and $T_a$ is the ambient temperature. If the power flows $\boldsymbol{F}_l\left( t^{\prime} \right)$ on edge $l$ at time $t^{\prime}$ is larger than its capacity $\boldsymbol{F}_{l}^{\max}$, edge $l$ is overloaded and the temperature will accumulate and approach to the critical temperature point $T_{l}^{*}=T_e\left( \boldsymbol{F}_{l}^{\max} \right)$. The time when edge $l$ approaches to its critical temperature point $T_{l}^{*}$ is,
\begin{equation}
	t_{l}^{*}=-\frac{1}{\mu}\ln \frac{T_{l}^{*}-T_e\left( \boldsymbol{F}_l\left( t^{\prime} \right) \right)}{T_l\left( t^{\prime} \right) -T_e\left( \boldsymbol{F}_l\left( t^{\prime} \right) \right)} \label{eq6}
\end{equation}
Among all the overloaded edges, the one that approaches the critical temperature point first will be removed from the subnetwork. Then we go back to Step 2.  If there is no overloaded edge, the cascading failure stops.
\end{itemize}
It is worthwhile emphasizing that the time in the cascading failure model according to the DC power flow has physics meaning determined by how long it takes for edges to reach the critical temperature point in Eq. \ref{eq6}. 
\end{appendices}

\section*{References}
\bibliography{iopart-num}

\end{document}